\documentclass[aps,prl,twocolumn,superscriptaddress]{revtex4-1}
\usepackage{graphicx}
\usepackage{amsmath}
\usepackage{amssymb}
\usepackage{psfrag}
\usepackage{natbib}
\usepackage{color,xcolor}
\begin{document}
\title{Interplay of coupling and common noise at the transition to synchrony in
oscillator populations}
\author{Anastasiya V. Pimenova}
\affiliation{Institute of Continuous Media Mechanics, UB RAS,
 Perm 614013, Russia}
\author{Denis S.\ Goldobin}
\affiliation{Institute of Continuous Media Mechanics, UB RAS,
 Perm 614013, Russia}
\affiliation{Department of Theoretical Physics, Perm State University,
 Perm 614990, Russia}
\author{Michael Rosenblum}
\affiliation{Institute for Physics and Astronomy,
University of Potsdam, Karl-Liebknecht-Str. 24/25, 14476 Potsdam-Golm, Germany}
\author{Arkady Pikovsky}
\affiliation{Institute for Physics and Astronomy,
University of Potsdam, Karl-Liebknecht-Str. 24/25, 14476 Potsdam-Golm, Germany}
\affiliation{Department of Control Theory, Nizhny Novgorod State University,
Gagarin Av. 23, 606950, Nizhny Novgorod, Russia}
\begin{abstract}
We consider a population of globally coupled oscillators driven by common noise. By applying
the Ott-Antonsen
ansatz and by averaging over the fast oscillations, we obtain analytically
tractable equations for
the noisy evolution of the order parameter. While noise always tends to
synchronize the oscillators, the
coupling can act against synchrony if it is repulsive. For identical
oscillators, the fully synchronous state
remains stable for  small enough repulsive coupling; moreover it is
an absorbing state
which always wins over the asynchronous regime. For oscillators
with a distribution of
natural frequencies, we report on a counter-intuitive effect of
dispersion of the oscillators frequencies at
synchrony.
\end{abstract}
\maketitle
%==============================================================================%
%\section{Introduction}
Synchronization effects in ensembles of coupled oscillators are
relevant for various physical systems,
such as coupled lasers, spin-torque oscillators, and Josephson
junctions~\cite{Nixon_etal-13,*Grollier-Cros-Fert-06,*Cawthorne_etal-99}, as well as for diverse
natural phenomena in life
sciences~\cite{Richard-Bakker-Teusink-Van-Dam-Westerhoff-96,*Prindle_etal-12}, and even for many
social systems~\cite{Neda_etal-00,*Eckhardt_et_al-07}. Synchronization
caused by an attractive mean-field coupling, studied in pioneering works by
Winfree and Kuramoto~\cite{Winfree-67,*Kuramoto-75},
allows a two-fold characterization. On one hand, the synchronization transition
can be described via
the appearance of a macroscopic mean field, amplitude of which often serves as the order parameter
of the transition.
On the other hand, synchronization can be characterized via an adjustment of
the frequencies of the
oscillators
in the ensemble (e. g., N. Wiener described synchronization~\cite{Wiener-65} as a ``phenomenon of
the pulling together of frequencies'').
There is also a nontrivial way to synchronize oscillators without coupling by acting
on them with
a common external noise~\cite{Pikovsky-84,*Pikovsky-84a,*Goldobin-Pikovsky-05b}.
Remarkably, common noise synchronizes
oscillators
in the first meaning only.
So, an ensemble of identical uncoupled oscillators under common noise
forms a perfect cluster where all the states
coincide and the value of the order parameter is the maximal possible.
The phases of slightly different oscillators also form a cluster. Their frequencies
are however not adjusted: their difference
is preserved under
common noise.

In this Letter we study properties of synchronization
and of the behavior of the frequencies
if both coupling and common noise are present. Our theory generalizes previous studies of
noise-driven ensembles
without coupling~\cite{Braun-etal-12}. We demonstrate how the Ott-Antonsen ansatz,
valid in the thermodynamic limit
for coupled oscillators with a Lorenzian distribution of
natural frequencies, can be generalized to include a common noisy driving.
After averaging of the resulting
equations over the fast basic frequency of oscillations,
we get a tractable Langevin-type
dynamics of the order parameter. We discuss in detail a nontrivial
competition between the synchronizing action
of noise and the desynchronizing action of the repulsive coupling. For
nonidentical oscillators, where
complete synchrony is impossible, we derive stationary distribution
of the order parameter and describe a rather counter-intuitive dispersion
 of the frequencies at synchronization in presence of the repulsive coupling.

%\section{Basic equations}
We consider an ensemble of phase oscillators subject to a
common Gaussian white noise with intensity $\sigma^2$
and to a Kuramoto-type coupling with strength $\mu$ (the coupling
is attractive for $\mu>0$ and repulsive otherwise).  We consider
the ensemble in the thermodynamic limit,
suitable for the application
of the Ott-Antonsen theory~\cite{Ott-Antonsen-08}:
\begin{equation}
\begin{gathered}
\dot{\varphi}_\Omega=\Omega+\sigma\xi(t)\sin\varphi_\Omega+
\mu R\sin(\Phi-\varphi_\Omega)\;,\\\quad \langle \xi(t)\xi(t')
\rangle=2\delta(t-t')\;.
\end{gathered}
\label{eq:1}
\end{equation}
Here the mean field is defined as
\begin{equation}
Z=Re^{i\Phi}=\langle e^{i\varphi}\rangle=
\int_{-\infty}^\infty d\Omega\, g(\Omega)\int_0^{2\pi}d\varphi_\Omega\, e^{i\varphi_\Omega}
w(\varphi_\Omega,t) \;,
\label{eq:2}
\end{equation}
where $g(\Omega)$ is the distribution of the natural frequencies.
According to the Ott-Antonsen ansatz~\cite{Ott-Antonsen-08},
the distribution function of the phases at given $\Omega$ can be represented
as $w(\varphi_\Omega,t)=(2\pi)^{-1}(1+\sum_{k=1}^\infty
[(z_\Omega(t))^k e^{-ik\varphi_\Omega}+c.c])$
and the mean field $z_\Omega(t)$ of a subpopulation with frequency $\Omega$
 obeys the equation
\begin{equation}
\dot z_\Omega=i\Omega z_\Omega+0.5[\mu Z-\sigma\xi(t)-
(\mu Z^*-\sigma \xi(t)) z^2_\Omega]\;.
\label{eq:3}
\end{equation}
For a Lorentzian distribution of frequencies
$g(\Omega)=\gamma[\pi(\gamma^2+(\Omega-\Omega_0)^2)]^{-1}$, the
integral in~\eqref{eq:2} can be calculated by virtue of the
residual theorem, under assumption of analyticity of $z_\Omega$ in the upper half-plane,
$
Z=\int_{-\infty}^\infty d\Omega g(\Omega) z_\Omega=z_{\Omega_0+i\gamma}
$.
As a result one obtains a closed equation for the
mean field $Z$ for coupled
non-identical oscillators under common noise:
\begin{equation}
\dot Z=i\Omega_0 Z-\gamma Z+0.5[\mu Z(1-|Z|^2)-\sigma(1-Z^2)\xi(t)]\;.
\label{eq:5}
\end{equation}
It contains four parameters: the basic frequency $\Omega_0$ (which, in
contradistinction to the usual
Kuramoto model, cannot be simply shifted to zero, because the noise term breaks
the frequency-shift invariance), the
noise intensity $\sigma^2$, the coupling constant $\mu$, and the width of the distribution
of natural frequencies $\gamma$.

For an analytical treatment below, it is convenient to use the real-valued variables
$(J,\Phi)$, where $J=R^2/(1-R^2)$
is the order parameter characterizing the level of synchrony (closeness of the phases of
oscillators in the ensemble):
for $J=0$ the mean field amplitude $R=\sqrt{J/(1+J)}$ vanishes, while the
full synchrony with $J=\infty$ corresponds to $R=1$. Equations for these variables read
\begin{equation}
\begin{aligned}
\dot J&=\mu J-2\gamma J(1+J)-\sigma\xi(t)\sqrt{J(1+J)}\cos\Phi\;,\\
\dot \Phi&=\Omega_0+\sigma \xi(t) (J+1/2)[J(1+J)]^{-1/2}\sin\Phi\;,
\end{aligned}
\label{eq:6}
\end{equation}
and are complemented with the
equation for the phase,
relative to that of the mean field, $\theta_\omega=\varphi_\Omega-\Phi$:
\begin{equation}
\begin{gathered}
\dot\theta_\omega=\omega-\mu\sqrt{J/(1+J)}\sin\theta_\omega+\\
\sigma\xi(t)[\sin(\Phi+\theta_\omega)-(J+1/2)[J(1+J)]^{-1/2}\sin\Phi]\;.
\end{gathered}
\label{eq:7}
\end{equation}
Here $\omega=\Omega-\Omega_0$ is the deviation of the natural frequency
from the ensemble mean one. For the sake
of simplicity of notations we omit index $\omega$ below.

%\section{Averaging}
As the first step, we employ the natural condition that
the basic frequency of oscillations $\Omega_0$
is much larger than the parameters $\mu,\gamma,\sigma^2$
(which all have dimension of inverse time).
This suggests to average over the fast rotating phase $\Phi$. One
writes the Fokker-Planck equation corresponding to the
Langevin equations~(\ref{eq:6},\ref{eq:7}), and by virtue of the
multiple scales expansion obtains in the leading order in the small
parameters $\mu,\gamma,\sigma^2$ the
following equation for the probability density $w(J,\theta,t)$:
\begin{equation}
\begin{gathered}
\frac{\partial w}{\partial t}+\frac{\partial}{\partial J}
([\mu J-2\gamma J(1+J)+\frac{\sigma^2}{2}(J+1/2)]w)\\
+\frac{\partial}{\partial\theta}\left(\left[\omega-\mu\sqrt{\frac{J}{1+J}}\sin\theta-
\frac{\sigma^2(J+1/2)}{2\sqrt{J(1+J)}}\right]w\right)\\
-\sigma^2 \hat Q_{J,\theta}^2 w-\sigma^2 \hat Q_\theta^2 w=0\;.
\end{gathered}
\label{eq:8}
\end{equation}
Here we defined the operators
\begin{equation}
\begin{aligned}
 \hat{Q}_{J,\theta}(\cdot)&\equiv
 \frac{\partial}{\partial J}\left(\sqrt{\frac{J(1+J)}{2}}\,(\cdot)\right)
 -\frac{\partial}{\partial\theta}\left(\frac{\sin{\theta}}{\sqrt{2}}\,(\cdot)\right),\\
\hat{Q}_\theta(\cdot)&\equiv
 \frac{\partial}{\partial\theta}\left(\left(\frac{\cos{\theta}}{\sqrt{2}}
 -\frac{J+1/2}{\sqrt{2J(1+J)}}\right)
 (\cdot)\right).
 \end{aligned}
\label{eq:9}
\end{equation}
The Fokker-Planck equation~(\ref{eq:8},\ref{eq:9}) is equivalent to the
following system of stochastic
Langevin equations which can be interpreted as Eqs.~(\ref{eq:6},\ref{eq:7})  averaged over
the fast oscillations with frequency $\Omega_0$:
\begin{align}
\dot J&=\mu J-2\gamma J(1+J)+
\frac{\sigma^2}{2}(J+1/2)-\sigma\sqrt{\frac{(1+J) J}{2}}\zeta_1(t)\;,
\label{eq:10-1}\\
\dot\theta&=\omega-\mu\sqrt{\frac{J}{1+J}}\sin\theta-
\frac{\sigma^2}{4}\frac{(J+1/2)}{\sqrt{J(1+J)}}\sin\theta\nonumber\\&
+
\frac{\sigma}{\sqrt{2}}
\sin\theta\zeta_1(t)+\frac{\sigma}{\sqrt{2}}\left(\cos\theta-\frac{(J+1/2)}{\sqrt{J(1+J)}}\right)\zeta_2(t)\;.
\label{eq:10-2}
\end{align}
The original noise $\xi(t)$ generates two effective independent
noise terms $\zeta_1(t)$ and $\zeta_2(t)$, which are Gaussian and
delta-correlated, $\langle\zeta_n(t)\zeta_l(t+t')\rangle=2\delta_{n,l}\delta(t')$\,,
because the signals $\xi(t)\cos{\varOmega_0t}$ and $\xi(t)\sin{\varOmega_0t}$ are
uncorrelated on time scales that are large compared to $2\pi/\varOmega_0$. The derived equations
contain four parameters $\mu,\gamma,\sigma^2,\omega$, and the properties of the stationary
solutions depend on $\mu/\sigma^2$, $\gamma/\sigma^2$, and $\omega/\sigma^2$ only.

Our first goal is to characterize the statistics of the order parameter $J$.
One can see that, as it should be for any
global coupling setup,
the system~(\ref{eq:10-1},\ref{eq:10-2}) is a skew one, where the dynamics of the
order parameter affects that of the phases,
but not vice versa. Thus one obtains a closed Fokker-Planck equation (the corresponding
Langevin equation is ~\eqref{eq:10-1}) for the
distribution of the order parameter
\begin{equation}
\begin{gathered}
\frac{\partial W(J,t)}{\partial t}+\frac{\partial}{\partial J}
([\mu J-2\gamma J(1+J)+\frac{\sigma^2}{2}(J+1/2)]W(J,t))\\=\frac{\sigma^2}{2}
\frac{\partial}{\partial J}\sqrt{J(1+J)}\frac{\partial}{\partial J}\sqrt{J(1+J)} W(J,t)\;.
\end{gathered}
\label{eq:fpJ}
\end{equation}

%\section{Identical oscillations}

We start by considering the case of identical oscillators $\gamma=0$. Here,
the analysis of states close to full synchrony
$J\to\infty$ is simple, as $\ln J$ performs a biased random walk:
\begin{equation}
\frac{d}{dt}\ln J=\mu+\frac{\sigma^2}{2}+\frac{\sigma}{\sqrt{2}}\eta_1(t)\;.
\label{eq:11}
\end{equation}
The quantity $\lambda=-\mu-\frac{\sigma^2}{2}$ is nothing else as the
Lyapunov exponent determining stability
of the full synchrony, the latter is stable if $\lambda<0$, i.e. if $\mu>-\sigma^2/2$.
Thus, the small enough repulsive coupling between the
oscillators does not break stability of the full synchrony. Another important state
is that of full asynchrony, $J=0$. One can see
however from Eq.~\eqref{eq:10-1} that this state is not invariant in presence of noise.

In fact, here we meet a nontrivial
situation where  the states of full asynchrony ($J=0$) and of full synchrony
($J=\infty$) are differently driven by noise.
For the asynchronous state the driving is additive, therefore this state is not
invariant and the order parameter
experience fluctuations close to $J=0$, even if this state is stable (i.e. for repulsive
coupling $\mu<0$). In contradistinction, the noise
is acting on the fully synchronous state in a multiplicative way, so that if this state
is stable, noise does not kick the system out of it.
Thus, the stable ($\lambda<0$) fully synchronous state is an absorbing one.
This means that also for a
slightly repulsive coupling $-\sigma/2<\mu<0$, the asynchronous state $J\approx 0$,
although stable without noise, does
not survive the competition with the fully synchronous state $J=\infty$ which is the
global attractor.

In this ``bistable'' situation the nontrivial statistical characteristics is the
 mean first passage time $T(0,\tilde J)$ for the stochastic
process ()\ref{eq:10-1},\ref{eq:fpJ}), from asynchrony
$J(0)=0$ to synchrony $\tilde J\gg 1$ (here
a cutoff is needed, because the
approach to the full synchrony $J=\infty$ is exponential,
formally the time to achieve it
is infinite). The expression for $T$ can be found via the standard
first-passage time theory for one-dimensional stochastic
processes~\cite{Gardiner-96}:
\begin{equation}
\begin{gathered}
T=\frac{2}{2\mu+3\sigma^2}\int_0^{\tilde J}
\frac{1-(1+z)^{2\mu\sigma^{-2}+3}}{z}dz\;.
\end{gathered}
\label{eq:12}
\end{equation}
Depending on the value of $\mu/\sigma^2$, this time
changes from a logarithmically large one $\sim\log\tilde J$
for
$2\mu\sigma^{-2}+3>0$, to a time diverging as a power law of $J$ for
$2\mu\sigma^{-2}+3<0$.
%\section{Nonidentical oscillators}

For nonidentical oscillators, $\gamma>0$, the fully synchronous state does not exist.
In this situation the order
parameter $J$ fluctuates with the stationary distribution, which can be
straightforwardly found from \eqref{eq:fpJ}:
\begin{equation}
\begin{gathered}
W(J;\gamma,\mu,\sigma^2)=\frac{ (1+J)^{2\mu\sigma^{-2}}
\exp[-4\gamma\sigma^{-2}(1+J)]}
{(4\gamma\sigma^{-2})^{1+2\mu\sigma^{-2}}\Gamma(2\mu\sigma^{-2}+1,4\gamma\sigma^{-2})}\;,\end{gathered}
\label{eq:15}
\end{equation}
where $\Gamma(m,x)$ is the upper incomplete Gamma function. The average value of the
order parameter is
\begin{equation}
\langle J\rangle=
\frac{1+2\mu\sigma^{-2}}{4\gamma\sigma^{-2}}-1+
\frac{\exp[-\frac{4\gamma}{\sigma^2}][\frac{4\gamma}{\sigma^2}]^{2\mu\sigma^{-2}}}
{\Gamma(1+2\mu/\sigma^2,4\gamma\sigma^{-2})}
\label{eq:avJ}
\end{equation}

\begin{figure}[tb]
\centering
\psfrag{xlab}[cc][cc]{$\mu/\sigma^2$}
\psfrag{ylab}[cc][cc]{$\langle J\rangle$}
\includegraphics[width=\columnwidth]{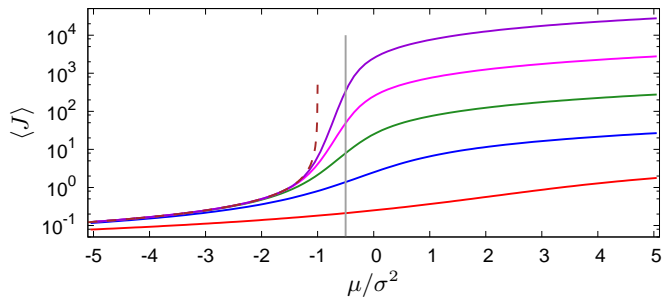}
\caption{(color online) Values of $\langle J\rangle$ for different $\gamma/\sigma^2$ as functions
of $\mu/\sigma^2$. From top to bottom:  $\gamma/\sigma^2=10^{-4},\;10^{-3},\;10^{-2},\; 10^{-1},\;1$.
Brown dashed line corresponds to the system of identical oscillators $\gamma=0$.
Vertical grey line shows the border
of stability of the fully synchronous state for $\gamma=0$.}
\label{fig:avJ}
\end{figure}

These expressions are valid for any $\gamma>0$, however
the limit $\gamma\to 0$ is singular:
a normalizable distribution for $J$ at $\gamma=0$
\begin{equation}
W(J;0,\mu,\sigma^2)=-(2\mu\sigma^{-2}+1)(1+J)^{2\mu\sigma^{-2}}
\label{eq:13}
\end{equation}
exists only
if the synchronous state is unstable, i.e. $\mu<-\sigma^2/2$,
and the average $\langle J\rangle=-(2\mu\sigma^{-2}+2)^{-1}$ is finite only if
$\mu<-\sigma^{-2}$.
We present the dependencies of $\langle J\rangle$ on the parameters
of the problem
in Fig.~\ref{fig:avJ}.

\begin{figure}[tb]
\centering
\psfrag{xlab}[cc][cc]{$\omega$}
\psfrag{ylab}[cc][cc]{$\nu$}
\includegraphics[width=\columnwidth]{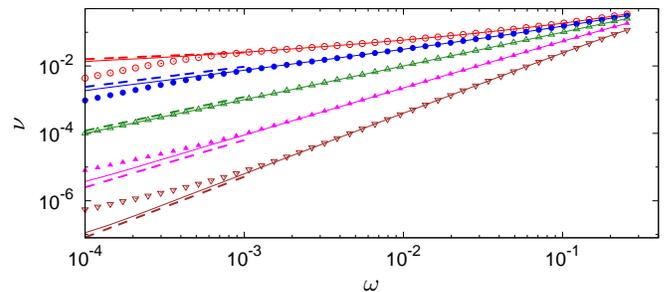}
\caption{(color online) Observed frequencies $\nu$ vs natural frequencies $\omega$, obtained
from~\eqref{eq:18}. We use here the continuous fraction expansion of the Fourier
representation of $w(\theta)$, following~\cite{Risken-89}. Solid lines: solutions for $J=\infty$,
markers: solutions for $\langle J\rangle=10$. From top to bottom:
$\mu/\sigma^2=-0.4,\;-0.2,\;0,\;0.2,\;0.4$. Dashed lines have slopes $1+2\mu/\sigma^2$.
}
\label{fig:avom}
\end{figure}

For nonidentical oscillators we face a new problem of the behavior of the frequencies. The skew
Langevin Eqs.~(\ref{eq:10-1},\ref{eq:10-2}) appear to be analytically solvable only if we make another 
approximation: We neglect fluctuations of the order parameter
(i.e. we assume $J\approx const$, for large $J$ this agrees with numerics)
in the equations for the phases. In this approximation
we obtain from Eq.~\eqref{eq:10-2} a closed Langevin equation for the phase dynamics:
\begin{equation}
\begin{gathered}
\dot\theta=\omega-\mu b\sin\theta-\frac{\sigma^2}{2}c\sin\theta\\
-\frac{\sigma}{\sqrt{2}}\sin\theta \zeta_1(t)+\frac{\sigma}{\sqrt{2}}
(\cos\theta-c)\zeta_2(t)\;,
\end{gathered}
\label{eq:16}
\end{equation}
where we  denote
$
b= \sqrt{J/(1+ J )}$, $c=(J+1/2)/
{\sqrt{ J (1+ J )}}
$.
The stationary solution $w(\theta)$ of the corresponding Fokker-Planck equation with
a constant flux $j=(2\pi)^{-1}\langle \dot\theta\rangle$ obeys
\begin{equation}
(\omega-\mu b\sin\theta)w-\frac{\sigma^2}{2}\frac{d}{d\theta}(1-2c\cos\theta+c^2)w =j\;.
\label{eq:17}
\end{equation}
Solution of this equation reads
\begin{equation}
\begin{aligned}
w(x)&=\frac{C}{1-2c\cos\theta+c^2}\int_x^{x+2\pi}\exp[V(y)-V(x)]dy\;,\\
V(x)&=
-\frac{4\omega}{\sigma^2(c^2-1)}
\arctan \left(\frac{1+c}{c-1}\tan\frac{x}{2}\right)\\&+
\frac{\mu b}{c\sigma^2}\ln(1+c^2-2c\cos x)\;,\\
C^{-1}&=\int_0^{2\pi}\frac{\int_x^{x+2\pi}\exp[V(y)-V(x)]dy}{1-2c\cos x+c^2}dx\;,\\
\langle\dot\theta\rangle&=\pi C\sigma^2[1-\exp[V(x+2\pi)-V(x)]]\\&=\pi C\sigma^2
\left(1-\exp\left[-\frac{4\pi \omega}{\sigma^2(c^2-1)}\right]\right)\;.
\end{aligned}
\label{eq:18}
\end{equation}
This rather lengthy exact solution can be simplified, for small $\omega$, to include
 the first-order terms $\sim \omega$ only.
Here the expression for $j$ reduces to $j\approx C\frac{2\pi\omega}{c^2-1}$, and
in the normalization factor $C$ we can set $\omega=0$:
\begin{equation}
\begin{gathered}
C^{-1}=\int_0^{2\pi}\frac{\int_x^{x+2\pi}
\left(\frac{1+c^2-2c\cos y}{1+c^2-2c\cos x}\right)^{\frac{\mu b}{c\sigma^2}}}
{1+c^2-2c\cos x}dx=\\
=\frac{4\pi^2}{c^2-1}\left(P_{\frac{\mu b}{c\sigma^2}}\left(\frac{c^2+1}{c^2-1}\right)\right)^2\;,
\end{gathered}
\label{eq:19}
\end{equation}
where $P_\lambda(x)$ is the Legendre function. A rather simple expression appears
for small $\mu$, where an expansion of the Legendre function can be used. The final
approximate formula for the observed frequencies of oscillators $\nu=\langle \dot \theta\rangle$
reads
\begin{equation}
\nu(\omega)=\langle\dot\theta\rangle=2\pi j\approx
\omega\left(1- 2\mu \frac{b}{c\sigma^2}\ln\frac{c^2}{c^2-1}\right)\;.
\label{eq:20}
\end{equation}
Noteworthy,  for uncoupled oscillators $\mu=0$ one obtains $\nu=\omega$.
This means that common noise
does not influence the average
frequencies. In the presence of coupling, the observed frequencies $\nu$
are pulled together if the coupling is attractive, $\mu>0$, and are pushed apart
if the  coupling is repulsive, $\mu<0$. The effect depends on the level of synchrony,
characterized by the value of the order parameter
$J$. In fact, the limit $J\to\infty$ is singular as here $c\to 1$; as we show in Fig.~\ref{fig:avom},
in this limit the dependence  $\nu(\omega)$ is not linear, but a power law one.

Formula \eqref{eq:20} describes, in an approximate way, the main nontrivial
effect that appears due to combined action of common noise and mean-field coupling
on the ensemble of nonidentical oscillators. We first remind what happens to the frequencies
in the absence of the common noise, i.e. for the standard Kuramoto model. In this case
there is a critical value of the coupling constant, beyond which the order parameter is non-zero.
In this synchronized state the frequencies are pulled together; moreover there appears
a cluster of oscillators that have equal frequencies, the size of this cluster grows with
the coupling constant. Below the critical coupling strength, the order parameter vanishes, so that there
is no any effect on the frequencies of the oscillators, and they remain the natural ones.

Common noise additionally influences the order parameter,
which is non-vanishing and even large also  when the mean-field
coupling is repulsive (cf. Fig.~\ref{fig:avJ}). This leads to a surprising
state of synchronization
with dispersion of the frequencies: synchrony (in the sense of a large value order parameter)
is in this case maintained by the common noise, while the repulsive coupling is responsible
for the scattering of frequencies.

As this effect is notable, we characterize it below numerically on different levels.
First, in Fig.~\ref{fig:avom} we show the solutions~\eqref{eq:18} for $J=\infty$ (perfect
synchronization) and for a finite $J$. One can see that in the fully synchronous case
$J=\infty$ the repulsion of frequencies is not linear as in Eq.~\eqref{eq:20}, but follows
a power law $\nu\sim \omega^\alpha$, with an exponent that with high accuracy can be fitted
as $\alpha= 1+2\mu/\sigma^2$.

\begin{figure}[tb]
\centering
\psfrag{xlab}[cc][cc]{$\omega/\gamma$}
\psfrag{ylab}[cc][cc]{$(\nu-\omega)/\gamma$}
\includegraphics[width=\columnwidth]{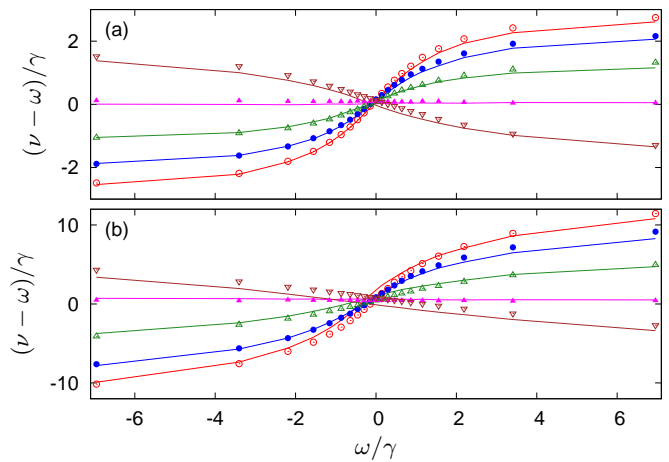}
\caption{Observed frequencies $\nu$ vs natural frequencies $\omega$, obtained
from~(\ref{eq:1},\ref{eq:5}). Parameters of simulations: $\Omega_0=100,\;\sigma=1$,
$\gamma=0.05$ (a) and $\gamma=0.01$ (b). Values of the coupling constant (from top to
bottom curves at the right side of the panels): $\mu=-0.6,\;-0.4,\;-0.2,\;0,\;0.2$.
Solid lines: simulations of the Ott-Antonsen equations~\eqref{eq:5}
valid in the thermodynamic limit. Markers: direct simulations of the population of
21 phase oscillators~\eqref{eq:1}.
}
\label{fig:oasim}
\end{figure}

Next, we illustrate in Fig.~\ref{fig:oasim} the effect of dispersion of the frequencies with the direct
simulation of
Langevin equations~(\ref{eq:1},\ref{eq:5})
describing the ensemble of coupled oscillators.  One  clearly sees
dispersion of the frequencies  for the repulsive
coupling and their concentration for the attractive coupling, both for the cases
of Ott-Antonsen equations~\eqref{eq:5} valid in the
thermodynamic limit, and for a finite population governed by~\eqref{eq:1}.

%\section{Conclusions}
In summary, in this Letter we have developed a theory for an ensemble of coupled
oscillators driven by common noise. In the thermodynamic limit, by adopting
the Ott-Antonsen ansatz and by averaging over the high basic frequency, we obtain
analytically tractable equations for the order parameter and find
the distribution of the order parameter in a closed form. As the common
noise always fosters synchrony of oscillators, nontrivial features appear if
the mean-field coupling acts in the opposite direction, i.e. is repulsive.
For identical oscillators this competition results in
the existence of the critical coupling strength $\mu_c=-\sigma^2/2$. For
$\mu>\mu_c$ the fully synchronous state where all the oscillators form a perfect
cluster is stable, while for $\mu<\mu_c$ it is not. Because, for vanishing noise,
the splay state with a uniform distribution of phases is stable for all negative
values of $\mu$, one could expect bistability for $\mu_c<\mu<0$. However,
bistability does not happen, because the noise acts differently at the two states
of interest: it is additive for the splay state with vanishing order parameter,
and is multiplicative for the fully synchronous state. The latter is thus an absorbing
state and the system never leaves it when the full synchrony is achieved.
Therefore for $\mu_c<\mu$ only the synchronous state is eventually observed, and the only
nontrivial question is how fast it is reached - the answer to this question is given by
Eq.~\eqref{eq:12}.

Another quite counter-intuitive effect of the competition between the common noise
and the coupling can be observed for non-identical oscillators.  The order
parameter is always non-vanishing in presence of common noise, and this leads
to dispersion of the frequencies - their distribution is wider than in the coupling-free case.
Here one should take into account that the common noise does not
directly adjust the frequencies,
although it pulls the phases together into a stochastic bunch.
In presence of an additional repulsive
coupling, the phases  in the bunch repel each other (although synchrony is preserved)
and as the
result their frequencies diverge.

%==============================================================================%

\acknowledgments
We thank P. Collet and M. Matias for useful discussions.
The work was supported by ITN COSMOS (funded by
the European Union’s Horizon 2020 research and innovation
programme under the Marie Sklodowska-Curie grant agreement No 642563).
Numerical part of this work was
supported by the Russian Science Foundation
(Project No. 14-12-00811).
The analytical calculations which led to Eqs.~(\ref{eq:6})--(\ref{eq:13}) was
supported by the Russian Science Foundation (Project No.\ 14-12-00090).
\def\cprime{$'$}
%

%\bibliography{nld-old,nld-current,%
%pap-ab,pap-ce,pap-fg,pap-hj,pap-kl,%
%pap-mn,pap-oq,pap-rs,pap-tz,%
%pik,books,n-stand,from_all_databases,%
%}
\end{document}